\documentclass[15pt]{article}

\usepackage{PRIMEarxiv}
\usepackage{natbib}
\usepackage[utf8]{inputenc} % allow utf-8 input
\usepackage[T1]{fontenc}    % use 8-bit T1 fonts
\usepackage{hyperref}       % hyperlinks
\usepackage{url}            % simple URL typesetting
\usepackage{booktabs}       % professional-quality tables
\usepackage{amsfonts}       % blackboard math symbols
\usepackage{nicefrac}       % compact symbols for 1/2, etc.
\usepackage{microtype}      % microtypography
\usepackage{lipsum}
\usepackage{fancyhdr}       % header
\usepackage{graphicx}       % graphics
\graphicspath{{media/}}     % organize your images and other figures under media/ folder
\usepackage{listings}
\usepackage{xcolor}
\usepackage{bm}
\usepackage{amsmath}

\definecolor{codegreen}{rgb}{0,0.6,0}
\definecolor{codegray}{rgb}{0.5,0.5,0.5}
\definecolor{codepurple}{rgb}{0.58,0,0.82}
\definecolor{backcolour}{rgb}{0.95,0.95,0.92}

\lstdefinestyle{mystyle}{
    backgroundcolor=\color{backcolour},   
    commentstyle=\color{codegreen},
    keywordstyle=\color{magenta},
    numberstyle=\tiny\color{codegray},
    stringstyle=\color{codepurple},
    basicstyle=\ttfamily\footnotesize,
    breakatwhitespace=false,         
    breaklines=true,                 
    captionpos=b,                    
    keepspaces=true,                 
    numbers=left,                    
    numbersep=5pt,                  
    showspaces=false,                
    showstringspaces=false,
    showtabs=false,                  
    tabsize=2
}
\author{Ying Zhang \\\text{UNC-Chapel Hill  }
   \And   John S. Preisser\\\text{UNC-Chapel Hill }
   \AND Fan Li\\Yale School of Public Health
   \And Elizabeth L. Turner\\Duke University
   \And Paul J. Rathouz\\University of Texas at Austin
   }

\title{\%CRTFASTGEEPWR: A SAS macro for power of generalized estimating equations analysis of multi-period cluster randomized trials with application to stepped wedge designs}

\begin{document}
\maketitle
\begin{abstract}
  Multi-period cluster randomized trials (CRTs) are increasingly used for the evaluation of interventions delivered at the group level. While generalized estimating equations (GEE) are commonly used to provide population-averaged inference in CRTs, there is a gap of general methods and statistical software tools for power calculation based on multi-parameter, within-cluster correlation structures suitable for multi-period CRTs that can accommodate both complete and incomplete designs. A computationally fast, non-simulation procedure for determining statistical power is described for the GEE analysis of complete and incomplete multi-period cluster randomized trials. The procedure is implemented via a SAS macro, \%CRTFASTGEEPWR, which is applicable to binary, count and continuous
responses and several correlation structures in multi-period CRTs. The SAS macro is illustrated in the power calculation of two complete and two incomplete stepped wedge cluster randomized trial scenarios under different specifications of marginal mean model and within-cluster correlation structure. The proposed GEE power method is quite general
as demonstrated in the SAS macro with numerous input options. The power
procedure and macro can also be used in the planning of parallel and crossover CRTs in
addition to cross-sectional and closed cohort stepped wedge trials.  
\end{abstract}

 
\section{Introduction} \label{sec:intro}
Cluster randomized trials (CRTs) are studies designed to evaluate interventions that operate at a group level, manipulate the physical or social environment, or cannot be delivered to individuals \citep{Murray2004}. Regarding the different schedules of recruiting participants, CRTs are classified with cross-sectional, closed-cohort and open-cohort designs \citep{Copas2015}. Cross-sectional designs recruit a unique set
of individuals in each period, whereas closed-cohort designs follow the same individuals in clusters
with repeated observations across periods. The open-cohort design, however, allows the attrition of
members from and addition of new members to an existing cohort in each period. On the other hand, there are different types of CRT designs, including parallel, crossover, and stepped-wedge designs. Among CRTs with outcomes measured in multiple periods, a stepped-wedge cluster randomized trial (SW-CRT) is a type of CRT such that clusters switch from control condition to treatment at randomly assigned time points  \citep{HusseyHughes2007}. Logistical and ethical considerations such as the need to deliver the intervention in stages and the desire to implement the intervention in all
clusters are factors involved in the choice to use a SW-CRT \citep{Turner2017}. SW-CRTs may be preferred over other designs because they may facilitate cluster recruitment or offer increased power over other cluster randomized designs even with a limited number of clusters \citep{Hemming2020}. Most study planning methods are for complete SW-CRTs where all clusters have outcome data in all periods.  However, incomplete stepped wedge designs are increasingly being deployed, whereby some cluster-periods do not record data due to logistical, resource, and patient-centered considerations \citep{kanza2019}. Specifically, researchers may choose not to collect data in a cohort design or enroll new participants in a cross-sectional design during some cluster-periods. \citet{Hemming2015a} described two types of incompleteness in stepped wedge designs, one involving implementation periods and the other staggered study entry or termination of clusters. 

Population-averaged models with GEE analysis have several advantages for the design and analysis of
CRTs \citep{Preisser2003}. In contrast to generalized linear mixed models, the intervention effect from a population-averaged model describes how the average response changes across the subsets of population defined by the treated and control cluster-periods. Additionally, because models for mean and correlation structures are separately specified, the interpretation of the marginal mean regression parameters remains the same regardless of the specification of working correlation model \citep{Preisser2008}.  The link
function is chosen to obtain inference on the target parameter of choice; for binary responses, the target parameters could be
the odds ratio via the logit link, the risk ratio via the log link, or the risk difference via the identify link. Another advantage in using GEE for CRTs is that the estimation of mean
model parameters is robust to misspecification of correlation structures in large samples. However, the specification of working independence correlation structure may result in efficiency loss that can be substantial when the cluster-period sizes are not all equal \citep{Tian2021}. Furthermore, an over-simplified exchangeable correlation structure may give inaccurate power calculations when there is correlation decay in multi-period CRTs \citep{Li2020,Kasza2019a}.
Thus, correlation structures informed by the study design are recommended for both study design and data analysis of stepped wedge and other multi-period CRTs. 

Because CRTs are usually less powerful than individually randomized trials, determination of the proper number and allocation of study participants is critically important. In the case of population-averaged models with GEE analysis, simple-to-use sample size formulae for continuous responses and non-simulation procedures for binary responses have recently been proposed for complete, cross-sectional and cohort SW-CRTs within the framework of GEE \citep{Li2018}. The methods extend earlier sample size formulae for GEE analysis of parallel-groups CRTs, including cross-sectional and cohort CRTs \citep{Preisser2003,Preisser2007} and multi-level CRTs \citep{Reboussin2012,Teerenstra2010,Wang2021}. Prior work \citep{Li2018,Li2020} has shown that the analytical power for marginal mean (e.g, intervention) parameters in complete SW-CRTs agrees well with simulated power based on GEE with finite-sample sandwich variance estimators for as few as eight clusters \citep{Li2018}. Those studies focus on the empirical performance of the analytical power rather than software tools to implement the method for different multi-period CRT designs.

This article implements a comprehensive, analytical power method for both complete and incomplete SW-CRTs. The proposed GEE power procedure is motivated by the Connect-Home trial, which uses an incomplete, cross-sectional stepped wedge design to test an intervention to improve outcomes for rehabilitation patients transitioning from skilled nursing facilities (SNFs) to home-based care \citep{Toles2021}. The primary component of the intervention is an individualized Transition Plan of Care that SNF staff create to support the patient and caregiver at home. The incomplete design with six SNFs (clusters) and four patients per cluster-period (360 patients total), shown in Figure 1, was chosen based on considerations of internal validity and power under restrictions placed by available resources and logistical considerations. The black and orange boxes represent cluster-periods where no patients are enrolled, giving an incomplete design. Staggered enrollment of SNFs (clusters) is used to initiate data collection in stages with limited research staff resulting in the black boxes. The orange boxes represent the implementation phase, where two months are needed to activate the intensive intervention through training nursing home and home health care staff. The Connect-Home study design is distinct in several aspects. First, the number of periods $(J=22)$ is much greater than the number of sequences $(S=6)$, as compared to the standard SW-CRT in which $J = S+1$ \citep{HusseyHughes2007,Hemming2015a,Li2018}. Next, the incompleteness of the design adds to the complexity of power calculation. 

Thus, the Connect-Home trial inspires an extension of the computationally fast, non-simulation procedures for determining sample size and statistical power for GEE analysis from complete SW-CRTs \citep{Li2018} to incomplete SW-CRTs. Specifically, we introduce a SAS\citep{SAS-STAT} macro \%CRTFASTGEEPWR as a computationally efficient, non-simulation based routine for determining the statistical power in multi-period CRTs by further accommodating incompleteness at the design stage. For its implementation, the SAS macro \%CRTFASTGEEPWR is developed to accommodate binary, count and continuous responses in stepped wedge and other multi-period CRTs with a collection of commonly-seen multilevel intra-cluster correlation structures and numerous options for planning complete and incomplete cross-sectional and cohort designs. To the best of our knowledge, \%CRTFASTGEEPWR is by far the most comprehensive SAS macro for power calculation in multi-period CRTs based on marginal models, and is distinct from existing computing software such as \text{R} package swdpwr \citep{Chen2022}, which currently does not allow for decaying correlation structures, incomplete designs, count outcomes and alternative marginal mean models beyond the average intervention effects model. 

The remainder of this article is organized as follows. Section 2 describes the population-averaged models of interest with special consideration of correlation structures suitable for cross-sectional and cohort multi-period CRTs. Section 3 summarizes the general power procedure for GEE analysis with the application to complete and incomplete stepped wedge designs. Section 4 presents the SAS macro details and four examples for complete and incomplete SW-CRTs.

\begin{figure}[h!]
\centering
\includegraphics[width = 14cm ]{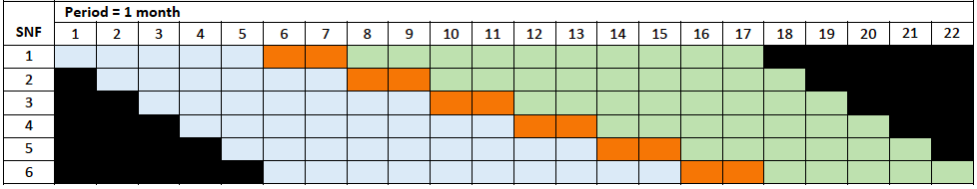}
\caption{The trial diagram of the Connect Home trial: the blue, orange and green cells denote control, implementation and intervention cluster-periods, respectively\label{fig1}}
\end{figure}

%% -- Manuscript ---------------------------------------------------------------

%% - In principle "as usual" again.
%% - When using equations (e.g., {equation}, {eqnarray}, {align}, etc.
%%   avoid empty lines before and after the equation (which would signal a new
%%   paragraph.
%% - When describing longer chunks of code that are _not_ meant for execution
%%   (e.g., a function synopsis or list of arguments), the environment {Code}
%%   is recommended. Alternatively, a plain {verbatim} can also be used.
%%   (For executed code see the next section.)

\section{GEE Analysis of multi-period CRTs} \label{sec:models}

A unifying population-averaged model framework is described below for the design and statistical analysis of multi-period CRTs. The following notations apply to both cross-sectional and cohort multi-period CRTs, where there are $J$ periods, $S$ sequences, $I$ clusters and $I_s$ clusters in sequence $s$, such that $I = \sum^S_{s=1}I_s$. Let $y_{ijk}$ denote the response of the $k$th individual from cluster $i$ during period $j$ for $i=1\ldots I, j = 1\ldots J_i$, and $k = 1\ldots N_{ij} $, noting that $J_i \leq J$ is the number of observed periods (i.e., with data collection) for cluster $i$, and $N_{ij}$ is the cluster-period size. Let $\mu_{ijk}$ denote the marginal mean response of $y_{ijk}$, which is related to the intervention effect $\delta$ and $j$th categorical period effect $\beta_j$ with link function $g(.)$ via the marginal mean model
\begin{equation}
    g(\mu_{ijk})=\beta_{j}+u_{ij} \delta
\end{equation}
For multi-period CRTs with fewer clusters than the number of periods ($I<J$), categorical period effects in the equation (1) can be replaced with linear period effects:
\begin{equation}
    g(\mu_{ijk})=\beta_{0}+\beta_{1}(t_{ij}-1)+u_{ij} \delta
\end{equation}
where $\beta_{0}$ is the intercept, and $\{t_{ij}: j=1,\ldots, J_i \}$ are integer-valued calendar periods from the study design such that $\beta_{1}$ is the increment in the mean response on the scale of the link function for a unit increase in calendar period, and $u_{ij}$ is the treatment status in cluster $i$ at period $j$. 

Three types of intervention effect models are implemented in the SAS macro, the widely used \textit{average intervention effects model} \citep{HusseyHughes2007,Hemming2015a,Li2018}, the \textit{incremental intervention effects model} \citep{Hughes2015}  and the \textit{extended incremental intervention effects model}. In the average intervention effects model, $u_{ij}$ is the period-specific treatment indicator (1=intervention; 0=control) for cluster $i$ and $\delta$ is the intervention effect, irrespective of time on treatment, on the link function scale. With the specification of $u_{ij}$, population-averaged models could be used for different types of multi-period CRTs. In multi-period CRTs with parallel designs, $u_{ij} = 0$ for all clusters under baseline period $j$, and $u_{ij} = 0, 1$ in subsequent post-baseline periods depending on the treatment status. In the case of complete SW-CRTs, clusters switch from control condition to intervention at different periods. Thus, $u_{ij}= 0$ in the control period, $j = 1, ..., b_i$ and $b_i $ is the total periods under the control condition in cluster $i$, whereas, in intervention periods, $u_{ij}= 1, j = b_i+1, ..., J_i$.
Conversely, the incremental intervention effects model assumes a gradual uptake of the intervention such that its effect depends on time-on-treatment. Assuming a complete SW-CRT is specified with the incremental intervention effects model, the treatment status is $u_{ij}=0$ in the control period and $u_{ij}= (j-b_i)/q, j = b_i+1, ..., J_i$ in the intervention period, where $q > 0$ is chosen to scale the intervention effect $\delta$ according to user specification. In the Connect-Home trial (Figure 1), $q=10$ so that $\delta$ is defined as the full intervention effect on the link function scale after 10 periods, which corresponds to the number of intervention periods for the first SNF. Finally, an extended incremental intervention effects model is considered for designs additionally having a maintenance phase after the active intervention phase with $q$ periods. In SW-CRTs with a maintenance 
phase, one research question relates to whether the patient benefit from the intervention as captured by the outcome is maintained after the active intervention period has ceased. The treatment status is $u_{ij}=(j-b_i)/q, j = b_i+1, ..., b_i+q$ in the active intervention periods and $u_{ij}=1$ in the maintenance phase, for $j >b_i+q$.

Specification of the marginal model is completed with the covariance structure of all individual responses in each cluster. Variance of the individual-level response is $\mbox{var}(y_{ijk}) = v_{ijk}\phi$ where $v_{ijk}$ is the variance function and $\phi$ is the dispersion parameter. For binary responses, $v_{ijk} = \mu_{ijk}(1-  \mu_{ijk})$ and $\phi=1$, while for continuous responses following typical normal model assumptions, $v_{ijk} =1$ and $\phi$ is the constant variance. For count outcomes with Poisson distribution, $v_{ijk} = \phi \mu_{ijk}$ and $\phi$ is the dispersion parameter. The macro assumes GEE is used to estimate the marginal mean parameters $\bm{\theta} = (\beta_1,...,\beta_p,\delta)$ as well as the working correlation matrix parameters; common multilevel correlation structures for multi-period CRTs will be presented in section 4.1 during the SAS macro description.

\section{Fast GEE power for multi-period CRTs}
\subsection{Overview of the fast GEE power method}

The two-sided test of the intervention effect $H_0$: $ \delta = 0$ vs $H_1$: $ \delta \neq 0$ is based upon the asymptotic normal distribution of $\sqrt{I}(\hat{\delta} - \delta)$ with mean zero and variance determined by the $(p,p)$-th element of cov$(\sqrt{I}(\hat{\bm{\theta}} - \bm{\theta}))$ , when $I$ is sufficiently large \citep{Li2018}. In turn, the Wald-test statistic $\hat{\delta}/\operatorname{var}(\hat{\delta})$ has an asymptotically standard normal distribution under the null hypothesis. Thus, power to detect an intervention effect of size $\delta\neq 0$ with a nominal type I error rate $\alpha$ is
 $\Phi\left(z_{\alpha / 2}+|\delta| / \sqrt{\operatorname{var}(\hat{\delta})}\right)$ where $\Phi(\cdot)$ is the standard normal cumulative distribution function and $z_{\alpha/2}$ is the normal quantile such that $\Phi(z_{\alpha/2}) =  \alpha/2$. However, for CRTs with a small number of clusters, the $t$-test is a good alternative with power modified as $\Phi_{t, I-p}\left(t_{\alpha / 2, I-p}+|\delta
| / \sqrt{\operatorname{var}(\hat{\delta})}\right)$ where $t_{\alpha/2, I-p}$ is the $\alpha/2\%$ quantile of the $t$-distribution with $I-p$ degrees of freedom. Typically, the degrees of freedom of the $t$-statistic in CRTs is set to $I - p$, where $p = \text{dim}(\bm{\theta})$ is the  number of estimated marginal mean model parameters in equation (1); some authors have used $I-2$, which is sometimes preferred for multi-period CRTs with fewer number of clusters than periods  \citep{Li2020,Ford2020}. The model-based variance of the intervention effect $\mbox{var}(\hat{\delta})$ in the determination of power is defined as the $(p,p)$-th element in the model-based covariance matrix $\text{cov}_{\text{MB}}(\hat{\boldsymbol{\theta}})$. We refer to this general analytical power method \citep{Rochon1998} as ``fast GEE power'' because it is a computationally fast power calculation procedure for CRTs with GEE analysis.

\subsection{Adaption of the fast GEE power for incomplete SW-CRTs}
The fast GEE power procedure has been previously investigated for complete SW-CRTs for the average intervention effects marginal mean model with categorical period effects \citep{Li2018}. Motivated by Connect-Home, we define a class of incomplete designs for which Connect-Home is an archetype allowing for implementation periods and/or staggered entry/termination. Let $b_{i0}$ and $b_{i1}$ denote the first and last calendar periods of data collection for cluster $i$ in the control condition ($b_{20}=2$ and $b_{21}=7$ for the second sequence, $s=2$, in Figure 1), such that there are $b_i = b_{i1} - b_{i0}+1$ total periods in the control condition; $q_{i0}$ and $q_{i1}$ as the first and last calendar periods of data collection for cluster $i$ in the intervention condition (e.g., $q_{20}=10$ and $q_{21}=18$, $s=2$) such that there are $q_i = q_{i1} - q_{i0}+1$ total periods in the intervention condition; and $c_i$ implementation periods occurring in calendar periods $b_{i1}+1, \ldots b_{i1}+c_i$ where $c_i = q_{i0} - b_{i1} -1$.

A key step in applying the fast GEE power computation to incomplete SW-CRTs is the generation of cluster-level design matrices, consisting of covariates in equation (1) or (2). For incomplete SW-CRTs, we specify a Design Pattern (DP) matrix by the notation of $\{b_{i0},b_{i1},q_{i0},q_{i1}, i=1,\ldots,I\}$ to represent the experimental design analogous to the power analysis of continuous responses in linear mixed models \citep{Hemming2015a}. Each element in the DP matrix corresponds to a representative cluster-period in the SW-CRT design with entries of 0 for control condition, 1 for intervention condition and 2 for cluster-periods in sequences without data collection. To illustrate the specification of DP matrix, an example of incomplete SW-CRT \citep{Hemming2015a} is used here with $S=2$ treatment sequences, $T=4$ periods, and an implementation period that occurs in period 2 for the first sequence and in period 3 for the second sequence. The incomplete design is specified by $(b_{10}, b_{11}, q_{10}, q_{11}) = (1,1,3,4)$ for $s=1$ and $(b_{20}, b_{21}, q_{20}, q_{21}) = (1,2,4,4)$ for $s=2$.  Then
 \begin{center}
        $\text{DP}=\left(\begin{array}{cccc}0& 2 &1 &1 \\ 0 & 0&2&1\end{array}\right).
        $
 \end{center}
The DP matrix serves to modify the design matrix obtained under the complete design to more accurately determine $\text{cov}_{\text{MB}}(\hat{\theta})$ and thus $var(\hat{\delta})$ in the power calculation for CRTs with incomplete design.
%% -- Illustrations ------------------------------------------------------------

\section{The SAS macro details} \label{sec:illustrations}
\subsection{Input arguments in the macro}
A SAS macro \%CRTFASTGEEPWR that implements the fast GEE power method is developed for multi-period CRTs with complete and incomplete designs and available at \newline
{\tt http://www.bios.unc.edu/$\sim$preisser/
personal/software.html\tt}. 

\begin{table}[ht!]
\begin{tabular}{p{6cm}p{2.0cm} p{8cm}}
\hline
 \textbf{Macro Variable*}&\textbf{Input} & \textbf{Description} \\\hline
\textbf{DESIGNPATTERN} & Variable &  A matrix with dimension S x T  of 0, 1 and 2,\\
&                        &with 0 representing control periods,1 standing for intervention periods, 2 presenting periods without data collection.\\
\textbf{CP\_SIZE\_MATRIX}        &Variable      & Number of cluster-period sizes, a matrix vector with dimension SxT\\
\textbf{M}                        &Variable      & Number of clusters in each sequence, a vector  with dimension Sx1  \\
\textbf{DIST}  & =BINARY  & The distribution for the outcomes \\
	      & =POISSON &  \\
	      & =NORMAL &  \\
LINK  & =LOGIT & Link function for the outcomes. For binary,\\
          & =LOG &   count, and continuous responses, the default \\
          & =IDENTITY &link is logit, log and identity, respectively \\

\textbf{PHI} & Variable & The dispersion parameter \\
\textbf{INTERVENTION\_EFFECT\_TYPE}  &=AVE          &Average intervention effects model\\
                             &=INC        &Incremental intervention effects model   \\ 
                             &=INC\_EX        &Extended incremental intervention effects model   \\ 
\textbf{PERIOD\_EFFECT\_TYPE}       &=CAT         & Categorical period effects model\\
                            &=LIN         & Linear period effects model \\
\textbf{DELTA}                      &Variable   & The parameters of intervention \\ 
\textbf{BETA\_PERIOD\_EFFECTS}      &Variable   & The parameters for period effects:\\
                           &            &A Tx1 vector for categorical period effect \\
		                   &            &A 2x1 vector for continuous period effect \\ 
		
\textbf{CORR\_TYPE} & =NE/ED &  NE:Nested Exchangeable, ED:Exponential decay,\\
& /BE/PD& BE:Block exchangeable, PD:Proportional decay  \\ 
\textbf{ALPHA0} &Variable & The within-period correlation in exponential decay and proportional decay correlation structure  \\ 
\textbf{R0}     &Variable  &Correlation decay rate over time in exponential decay and proportional decay correlation structure \\
\textbf{ALPHA1} &Variable  &The within-period correlation in nested exchangeable and block exchangeable correlation structure\\ 
\textbf{ALPHA2} &Variable    &The inter-period correlation in nested exchangeable and block exchangeable correlation structure\\
\textbf{ALPHA3} &Variable    &The within-subject correlation in block exchangeable correlation structure\\

MAX\_INTERVENTION\_PERIOD & Variable & The number of intervention periods to reach full intervention effects in incremental intervention \\
&    & effects model and extended incremental intervention effects model\\

ALPHA                 &Variable  & Significance level with two sided test, default at 0.05  \\ 
DF\_CHOICE &=1   & Degree of freedom method (df), df = I-p, p is the number of parameters in marginal mean model (at default) \\
          &=2   & df = I-2  \\
 \hline
\end{tabular}
\caption{Arguments in the SAS macro CRTFASTGEEPWR. *All required arguments are in boldface }
\end{table}

Table 1 provides required and optional arguments in the macro, which are classified into three aspects: describing the characteristics of the multi-period CRT, parameterizing the marginal mean model and choosing the working correlation structure. 

First, users are required to describe the characteristics of the multi-period CRT through the design pattern matrix, specified by $\textit{DESIGNPATTERN}$, containing the number of sequences and periods, numerical indicators for treatment status, and the incompleteness in the design. The SAS macro applies to cross-sectional and closed-cohort multi-period CRT designs but not to closed cohort designs. Specifically, the SAS macro allows a varying number of participants across cluster periods for cross-sectional designs through the specification of $\textit{CP\_SIZE\_MATRIX}$. For closed cohort designs, each column of $\textit{CP\_SIZE\_MATRIX}$ should be the same such that cluster-period sizes may vary across sequences but not within rows. The number of clusters in sequences are specified by a vector $\textit{M}$ to allow varying cluster numbers across sequences. 
 \newline
Marginal mean  model options include binary, count, and continuous responses with three link functions, specified by $\textit{DIST}$ and $\textit{LINK}$, respectively. Note that the default link function is the canonical link. Meanwhile, the categorical period effects model or the linear period effects model is selected by specifying $\textit{PERIOD\_EFFECT\_TYPE}$. The user also needs to choose one of three intervention effects models introduced in Section 2, specified by  $\textit{INTERVENTION\_EFFECT\_TYPE}$. 
For the incremental intervention effects model, $\textit{MAX\_INTERVENTION\_PERIOD}$ should be filled with the number of periods at which the full treatment effect $\delta$ is reached. For the extended incremental intervention effects model, the $\textit{MAX\_INTERVENTION\_PERIOD}$ is the number of periods under active intervention phase and there should be at least one maintenance period in each sequence. The intervention effect size and period effects at the scale of link function are all required with the choice of specific intervention and period effects model through $\textit{DELTA}$ and $\textit{BETA\_PERIOD\_EFFECTS }$. 

Four within-cluster correlation structures commonly used for multi-period CRTs are considered in the SAS macro (Table 2) through specification of $\textit{CORR\_TYPE}$ and corresponding intra-cluster correlations (ICCs). Specifically, there are two distinct correlation structures for each of the cohort and cross-sectional design. 
Each structure incorporates the usual ICC, which measures the correlation between responses from different individuals within the same cluster during the same period: $\text{corr}\left(y_{i j k}, y_{i j k^{\prime}}\right)=\alpha_{0}$ or $\alpha_{1}$, $k\neq k^{\prime}$. For cross-sectional designs, the $\textit{nested exchangeable}$ correlation structure additionally specifies a correlation parameter $\alpha_2$ for observation pairs collected from different periods. Alternatively, $\textit{exponential decay}$ assumes the between-period correlation between responses from different individuals within the same cluster in the $j_{th}$ and $j^{\prime}_{th}$ periods decays over time as $\alpha_0r_0^{|j-j^{\prime}|}$. For cohort designs, the $\textit{block exchangeable}$ correlation structure distinguishes between-period correlations for pairs of individuals, $\alpha_2,$ from a constant intra-individual correlation for repeated observations, $\alpha_3$  \citep{Li2018,Preisser2008}. On the other hand, the $\textit{proportional decay}$ correlation structure \citep{Li2020, Lefkopoulou1989} allows for correlation decay over time, where the intra-individual correlation $\text{corr}\left(y_{i j k}, y_{i j^{\prime} k}\right)=r_0^{|j-j^{\prime}|},  j \neq j^{\prime}$ has a first-order auto-regressive structure decay rate $r_0$ and the between-period correlation among responses from different individuals within the same cluster is $\alpha_0r_0^{|j-j^{\prime}|}, j \neq j^{\prime},  k\neq k^{\prime}$. Note that the nested exchangeable correlation for cross-sectional designs is a special case of block exchangeable correlation when $\alpha_2 = \alpha_3$.  Finally, the simple exchangeable correlation \citep{HusseyHughes2007} additionally specifies that within- and between-period correlations are equal $\alpha_1 = \alpha_2$. 
\begin{table}[ht!]
    \centering
\begin{tabular}{lllllll}
        \hline
        \text {Design } & \text {Correlation Structure} &\text{Label}& \text { $j = j^\prime $} & \text {$j \neq j^\prime$,$ k \neq k^\prime $ } & \text { $j \neq j^\prime$ ,$ k = k^\prime $} \\
        \hline {\text {Cross Sectional }} 
       & \text {Nested Exchangeable} &NE& $\alpha_{1}$ &$\alpha_{2}$&- \\

        &
        \text {Exponential Decay}&ED&
      $\alpha_0$  &$\alpha_0 r_0^{|j-j^{\prime}|}$&-
        \\
       \hline \text {Cohort }&  \text {Block Exchangeable}&BE&
        $\alpha_{1}$ &$\alpha_{2} $& $\alpha_{3}$\\
        & \text {Proportional Decay}&PD& $\alpha_0$  &$ \alpha_0r_0^{|j-j^{\prime}|}$
        & $r_0^{|j-j^{\prime}|}$
        \\
        \hline
\end{tabular}
           \caption{Intra-cluster correlation: $\operatorname{corr}\left(y_{i j k}, y_{i j^{\prime} k^{\prime}}\right)$ under different correlation structures for SW-CRTs (i=cluster, j=period, k=individual)}
\end{table}

The significance level for two-sided tests, specified by \textit{ALPHA} is optional with a default value of 0.05. In the degrees of freedom determination, the number of clusters minus the number of marginal mean parameters, $df = I-p$, is the default formula. Another degrees of freedom determination $df = I-2$ in \citet{Li2020} is also available, specified by $\textit{DF\_CHOICE}$.

There are some consistency checks in SAS macro \%CRTFASTGEEPWR to ensure parameters are reasonably specified for the power calculation. First, if there are cluster periods in the design pattern matrix with no data collection (i.e. 2 in \textit{DESIGNPATTERN}), the corresponding locations in the \textit{CP\_SIZE\_MATRIX} need to be 0. Second, \textit{BETA\_PERIOD\_EFFECTS} should match the selected period effects model. For the linear period effects model, there are two period effects parameters. While for the categorical period effects model, the number of period effects parameters is equal to the column size of the Design pattern matrix \textit{DESIGNPATTERN}. Finally, the marginal mean outcome $\mu_{ijk}$ needs to be within the rational range based on the specific outcome type, such as $\mu_{ijk}$ within (0,1) for binary outcomes and $\mu_{ijk}$ are non-negative for count outcomes. Specifically, for binary outcomes, the Fr\`echet bounds are also checked based on the specification of working correlation structures and the marginal means to ensure their compatibility \citep{Qaqish2003}.

\subsection*{4.2: Multi-period CRT examples of SAS macro \%CRTFASTGEEPWR }
In the section, we focus on illustrating the power calculation of two complete and two incomplete stepped wedge cluster randomized trials using the SAS macro \%CRTFASTGEEPWR, with different outcome types, specification of marginal mean models and correlation structures. 

The first example illustrates power calculation based on the Connect-Home trial design (Figure 1) with linear period effects for a continuous outcome, patient preparedness for home care (a scale with range 0 to 100) assessed 7 days 
after discharge from the SNF. There are 6 sequences with 22 periods in the study, having 7 periods without patient enrollment and 15 periods with patient enrollment in each cluster. In the power calculation, there is 1 cluster in each sequence and 360 subjects in the trial, in which 4 patients enrolled in each non-missing cluster period. In $\textit{CP\_SIZE\_MATRIX}$, 0 means no patients enrolled in the specific sequence and period, which corresponds to locations with 2 in the design pattern matrix. We assume the baseline patient preparedness score as $\beta_{0}=68$ and a gently increasing linear period effect such that $\beta_{1}=0.1$ for $J= 1,...,22$ with common variance $\phi =64$ (standard deviation = 8). The full effect size is reached at 10 months on intervention condition $(q=10)$ for the incremental intervention effects model with $\delta = 10$. ICCs are specified with $(\alpha_1, \alpha_2) = (0.1, 0.05)$ under nested exchangeable correlation structure to indicate a moderate within-cluster correlation for the cross-sectional design. The power is calculated using $z$-test with normal approximation and $t$-test with $df = 3$. From the results, the power using the z-test is much greater than the $t$-test. Simulation studies have shown that the $z$-test is too optimistic and tends to have an inflated test size in SW-CRTs with a small number of clusters \citep{Li2018}. Thus, we recommend calculating power with the $t$-test for the Connect Home study.

\begin{lstlisting}
%CRTFASTGEEPWR(alpha=0.05, m =%str(J(6,1,1)), corr_type = NE,alpha1 = 0.03, 
alpha2 = 0.015, intervention_effect_type=INC, delta = 10, period_effect_type=LIN,
max_intervention_period=10, beta_period_effects =%str({68,0.1}), dist = normal,
phi=64, 
CP_size_matrix = %str({4 4 4 4 4 0 0 4 4 4 4 4 4 4 4 4 4 0 0 0 0 0,
                       0 4 4 4 4 4 4 0 0 4 4 4 4 4 4 4 4 4 0 0 0 0,
                       0 0 4 4 4 4 4 4 4 0 0 4 4 4 4 4 4 4 4 0 0 0,
                       0 0 0 4 4 4 4 4 4 4 4 0 0 4 4 4 4 4 4 4 0 0,
                       0 0 0 0 4 4 4 4 4 4 4 4 4 0 0 4 4 4 4 4 4 0,
                       0 0 0 0 0 4 4 4 4 4 4 4 4 4 4 0 0 4 4 4 4 4}),
                       
DesignPattern  = %str({0 0 0 0 0 2 2 1 1 1 1 1 1 1 1 1 1 2 2 2 2 2,
                       2 0 0 0 0 0 0 2 2 1 1 1 1 1 1 1 1 1 2 2 2 2,
                       2 2 0 0 0 0 0 0 0 2 2 1 1 1 1 1 1 1 1 2 2 2,
                       2 2 2 0 0 0 0 0 0 0 0 2 2 1 1 1 1 1 1 1 2 2,
                       2 2 2 2 0 0 0 0 0 0 0 0 0 2 2 1 1 1 1 1 1 2,
                       2 2 2 2 2 0 0 0 0 0 0 0 0 0 0 2 2 1 1 1 1 1}));
\end{lstlisting}
\begin{lstlisting}
Output of Example 1:The fast GEE power of normal outcomes with nested exchangeable 
correlation structure and (alpha1,alpha2):(0.03, 0.015) under incremental
intervention effects model and delta = 10    

T   S  clusters   df   theta  totaln   Dist    Link     stddel    zpower    tpower
22  6     6        3    68     360    NORMAL   IDENTITY 3.9139    0.9746    0.7413
                        0.1                                                        
                        10   
\end{lstlisting}
The second example aims to calculate the power for a count outcome based on the Connect-Home trial design (Figure 1), which is the number of days of acute care use for patients within 60 days after discharge from the SNF. In this example, an average intervention effect model is specified with the same linear period effect model as in the first example. The design is identical to the first example but the number of clusters increases from 6 to 12 clusters to achieve enough power for the count outcome. For the parameters in the marginal mean model with log link, baseline number of acute care use is assumed to be 1.24 days with an slightly decreasing period effect over time, giving  $\beta_{0}  = \log(1.24) = 0.215$ and $\beta_{1}  = -0.01$ with the dispersion parameter $\phi = 1.2$. Assuming the intervention reduces the mean number of acute care days by 40\%, the intervention effect under the average intervention effects model is $\delta = \log(0.6) = -0.511$. ICCs are specified with $(\alpha_0, r_0) = (0.03, 0.8)$ under exponential decay correlation structure to indicate a small within-cluster correlation for the cross-sectional design. The power is calculated using the $z$-test with normal approximation and the $t$-test with $df = 9$. Compared to the first example that had only six clusters, power using the two tests are closer in values and exceed 79\%.

\begin{lstlisting}
%CRTFASTGEEPWR(alpha=0.05, m =%str(J(6,1,2)), corr_type = ED,alpha0 = 0.03,
R0 = 0.8, intervention_effect_type=AVE, delta = -0.511, period_effect_type=LIN,
beta_period_effects =%str({0.215,-0.01}) ,dist = poisson,phi=1.2, 
CP_size_matrix = %str({4 4 4 4 4 0 0 4 4 4 4 4 4 4 4 4 4 0 0 0 0 0,
                       0 4 4 4 4 4 4 0 0 4 4 4 4 4 4 4 4 4 0 0 0 0,
                       0 0 4 4 4 4 4 4 4 0 0 4 4 4 4 4 4 4 4 0 0 0,
                       0 0 0 4 4 4 4 4 4 4 4 0 0 4 4 4 4 4 4 4 0 0,
                       0 0 0 0 4 4 4 4 4 4 4 4 4 0 0 4 4 4 4 4 4 0,
                       0 0 0 0 0 4 4 4 4 4 4 4 4 4 4 0 0 4 4 4 4 4}),
    		            
DesignPattern =  %str({0 0 0 0 0 2 2 1 1 1 1 1 1 1 1 1 1 2 2 2 2 2,
                       2 0 0 0 0 0 0 2 2 1 1 1 1 1 1 1 1 1 2 2 2 2,
                       2 2 0 0 0 0 0 0 0 2 2 1 1 1 1 1 1 1 1 2 2 2,
                       2 2 2 0 0 0 0 0 0 0 0 2 2 1 1 1 1 1 1 1 2 2,
                       2 2 2 2 0 0 0 0 0 0 0 0 0 2 2 1 1 1 1 1 1 2,
                       2 2 2 2 2 0 0 0 0 0 0 0 0 0 0 2 2 1 1 1 1 1}));
\end{lstlisting}
\begin{lstlisting}{sas}
Output of Example 2: The fast GEE power of poisson outcomes with exponential decay
correlation structure and (alpha0,r0):(0.03, 0.8) under average intervention 
effects model and delta = -0.511

T   S  clusters  df   theta   totaln    Dist     Link    stddel   zpower   tpower
22  6    12       9   0.215    720     POISSON   LOG     3.1096   0.8749   0.7906
                      -0.01                                                     
                      -0.511                                                     

\end{lstlisting}

The third example illustrates power calculation for a cross-sectional SW-CRT to improve pre-operative decision-making, by the use of a patient-driven question prompt list intervention \citep{Taylor2017,Schwarze2020}. In the third example, 480 patients enrolled across
six periods are clustered within 40 surgeons who are randomized to transition from control (blue cells) to intervention condition (green cells) at one of five randomly assigned sequences (8 surgeons per sequence; Figure 2). We calculate the power for a binary primary outcome regarding whether the patient has a post-treatment regret. We assume a balanced and complete design for the study, 12 patients for each surgeon with two patients in each cluster period. In the marginal mean model for the binary outcome with logit link and average intervention effects model, the control is assumed to have 2.2 times the odds of reporting post treatment regret compared to the intervention group, given by $\delta= \log(1/2.2) = -0.789$. The average probability of post treatment regret at baseline is assumed to be 0.22, such that $\beta_0 = \log(0.22/0.78) = -1.266$ with an consistent increasing period effects $\beta_i = 0.01, i = 1,\ldots,5$. For the working correlation structure, we used the exponential decay correlation structure with ICCs $(\alpha_0, r_0) = (0.03, 0.8)$. Power using the $t$-test and the $z$-test both reach 80\% and are similar to one another considering the moderately large number of clusters. 
\begin{figure}[h!]
\centering
\includegraphics[width = 9cm ]{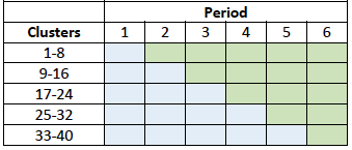}
\caption{The trial diagram of the decision-making trial: the study includes 40 clusters and each treatment sequence includes 8 clusters. The blue and green cells denote control and intervention cluster-periods, respectively\label{fig1}}
\end{figure}
\begin{lstlisting}{sas}
%CRTFASTGEEPWR(alpha=0.05, m =%str(J(5,1,8)), corr_type = ED, alpha0 = 0.03, 
R0 = 0.8, intervention_effect_type=AVE, delta =-0.789, period_effect_type=CAT, 
beta_period_effects =%str({-1.266,0.01,0.01,0.01,0.01,0.01}), dist = binary,phi=1,
CP_size_matrix =%str(J(5,6,2)),
DesignPattern =%str({ 0 1 1 1 1 1,
                      0 0 1 1 1 1,
                      0 0 0 1 1 1,
                      0 0 0 0 1 1,
                      0 0 0 0 0 1}));
\end{lstlisting}

\begin{lstlisting}{sas}
Output of Example 3: The fast GEE power of binary outcomes with exponential decay 
correlation structure and (alpha0,r0):(0.03, 0.8) under average intervention 
effects model and delta = -0.789   

T   S  clusters  df   theta    totaln   Dist    Link   stddel    zpower   tpower
6   5    40      33   -1.266    480     BINARY  LOGIT  2.9170    0.8307   0.8081
                       0.01                                                     
                       0.01                                                     
                       0.01                                                     
                       0.01                                                     
                       0.01                                                     
                      -0.789 
\end{lstlisting}
The fourth example is based on the Heart Health NOW study for which we assume a complete, stratified, SW-CRT evaluating the effect of primary care practice support on evidence-based cardiovascular disease (CVD) prevention, organizational change process measures, and patient outcomes, the latter captured by electronic health records (EHR) \citep{Weiner2015}. Medical practices are randomized to receive the intervention at one of three time points (steps) within two strata defined by high (the first three treatment sequences, Figure 3) or low (last three rows) readiness for change. After four quarters in the intervention phase (green boxes), each practice enters a maintenance phase (gray boxes) for two to six quarters depending upon the allocated treatment sequence. HHN was a quality improvement research project whereby the intervention of practice facilitation aimed to bring about enduring change in the patient-centered and organizational outcomes. We consider a combined binary outcome regarding whether there is hospitalization due to stroke, acute myocardial infarction, or angina for patients. We assume that there are 30 medical practices (clusters) in each sequence (practice cohort) with 100 patients enrolled in each cluster period; thus 1100 patients will be enrolled in each cluster. If the baseline probability of hospitalization is 0.05, making $\beta_0 = \log(0.05/0.95) = -2.944$ with an consistent decreasing period effects $\beta_1 = -0.01$. Under the extended incremental intervention effects model, the intervention effect is assumed to decrease the odds of hospitalization at the end of 4 quarters by 25\% under the active intervention condition, $\delta = \log(0.75) = -0.288$, maintaining the same effect size in the maintenance periods. The working correlation structure for the binary outcome is a nested exchangeable correlation structure with ICCs $(\alpha_1, \alpha_2) = (0.03, 0.015)$. Considering the large cluster size and number of clusters, powers is very similar under the $z$-test and $t$-test, reaching 78\%.
\begin{figure}[h!]
\centering
\includegraphics[width = 15cm ]{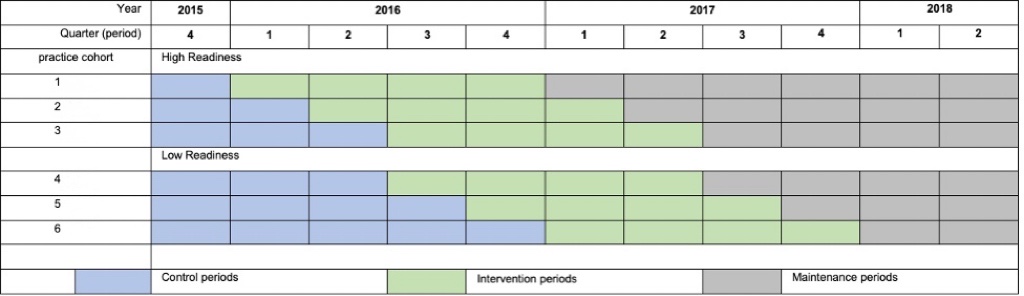}
\caption{The study design of the HHN study: the blue, green and grey cells denote control, active intervention and maintenance cluster-periods, respectively\label{fig1}}
\end{figure}
\begin{lstlisting}{sas}
%CRTFASTGEEPWR(alpha=0.05, m=%str(J(6,1,30)), corr_type=NE, alpha1=0.03, 
alpha2=0.015, intervention_effect_type=INC_EX, period_effect_type=LIN,
delta=-0.288, max_intervention_period=4,beta_period_effects=%str({-2.944,-0.01}),
dist = binary,phi=1, CP_size_matrix =%str(J(6,11,100)),
DesignPattern=%str({0 1 1 1 1 1 1 1 1 1 1,
                    0 0 1 1 1 1 1 1 1 1 1,
                    0 0 0 1 1 1 1 1 1 1 1,
                    0 0 0 1 1 1 1 1 1 1 1,
                    0 0 0 0 1 1 1 1 1 1 1,
                    0 0 0 0 0 1 1 1 1 1 1}));
\end{lstlisting}

\begin{lstlisting}{sas}
Output of Example 4: GEE power of binary outcomes with nested exchangeable 
correlation structure and (alpha1,alpha2):(0.03, 0.015) under extended incremental 
intervention effects model and delta = -0.288

T   S  clusters   df   theta   totaln   Dist     Link     stddel   zpower   tpower
11  6    180      177  -2.944  198000   BINARY   LOGIT    2.7477   0.7846   0.7801
                       -0.01                                                     
                       -0.288                                                     
 
\end{lstlisting}
In the appendix, we also describe the power calculation of multi-period parallel cluster randomized trials, as well as SAScodes to calculate powers under varying effect sizes using the SAS macro.

\section{Discussion}
This article proposes a fast GEE power method for binary, count, and continuous responses of complete and incomplete multi-period CRTs, including parallel-arm longitudinal CRTs, cluster randomized crossover trials and SW-CRTs. The fast GEE power approach was illustrated in the planning of four complete and incomplete cross-sectional stepped wedge designs for binary, count and continuous outcomes, with different correlation structures. The SAS macro \%CRTFASTGEEPWR is novel in several aspects. Through specification of the Design Pattern matrix in the spirit of Hemming et al. \citep{Hemming2015a} and Rochon, \citep{Rochon1998}, the general GEE power method in the SAS macro is implemented for complete and incomplete multi-period CRTs. We have also considered four multilevel correlation structures proposed in the recent literature for multi-period CRTs. To our knowledge, there is no other single statistical software that unifies the four multilevel correlation structures in an approach for designing cross-sectional and cohort CRTs. This proposed software for power of multi-period CRTs based on marginal models fills a gap by adding to the literature of power calculators based on mixed models. \citep{Hughes2015,Hemming2020a}

The accuracy of the fast GEE power method based on the GEE model-based variance matrix has been validated in simulation studies based on GEE with the bias-corrected variance estimator proposed by Kauerman and Carroll. \citep{Li2021,Kauermann2001} For researchers interested in GEE analysis with bias-corrected variance estimators for CRTs, the SAS macro $\textbf{GEECORR}$ by Shing and Preisser \citep{SHING2021106276} is available for binary responses. A SAS macro $\textbf{GEEMAEE}$ developed by the authors for binary, count, and continuous outcomes additionally applies bias-corrections in the estimation of ICCs and their variance estimators, which is desired for reporting ICC parameters as recommended by the CONSORT statement for stepped wedge trials \citep{Hemmingk1614}. $\textbf{GEECORR}$ and $\textbf{GEEMAEE}$ are all available at 
{\tt http://www.bios.unc.edu/$\sim$preisser/
personal/software.html\tt}.

\section*{Acknowledgments}
The Connect-Home trial was funded by the National Institute of Nursing Research of the National Institutes of Health under award number 1R01NR017636-01. The content is solely the responsibility of the authors and does not necessarily represent the official views of the National Institutes of Health.

Research in this article was funded through a Patient-Centered Outcomes Research Institute\textsuperscript{\textregistered} (PCORI\textsuperscript{\textregistered} Award ME-2019C1-16196).The statements presented in this article are solely the responsibility of the authors and do not necessarily represent the views of PCORI\textsuperscript{\textregistered}, its Board of Governors or Methodology Committee.

%% -- Bibliography -------------------------------------------------------------
%% - References need to be provided in a .bib BibTeX database.
%% - All references should be made with \cite, \citet, \citep, \citealp etc.
%%   (and never hard-coded). See the FAQ for details.
%% - JSS-specific markup (\text, \pkg, \code) should be used in the .bib.
%% - Titles in the .bib should be in title case.
%% - DOIs should be included where available.

\bibliography{citation}
\bibliographystyle{plainnat}

%% -- Appendix (if any) --------------------------------------------------------
%% - After the bibliography with page break.
%% - With proper section titles and _not_ just "Appendix".

\newpage
\section*{Appendix}
\begin{appendix}
In the appendix, we will illustrate the use of SAS macro \%CRTFASTGEEPWR for a parallel cluster trial, based on the Enforcing Underage Drinking Laws (EUDL) Program \citep{Preisser2003}. The EDUL program funded interventions at the community level to enforce laws related to alcohol use by underage person to reduce the underage drinking. Moreover, the study used a non-randomized trial design because the intervention communities were selected by the administrative units in states. The control communities were selected by the propensity score method to match the intervention communities based on US census data. There are three periods: one baseline assessment and two follow-up assessments for participants in the communities participating in the EUDL program. We will use the design of the EUDL study to calculate the power under the assumption that all confounded covariates were balanced in control and intervention groups. The main outcome is the binary outcome of self-reported last 30-day alcohol use for an underage person. We assume there are 40 clusters in total with 20 clusters per intervention group and 30 participants enrolled in each cluster-period. Assuming the baseline probability of self-reported last 30-day alcohol use for an underage person is 0.6, we set $\beta_0 = \log(0.6/0.4) = 0.405$ with an consistent decreasing period effects $\beta_1 = -0.01$. Under the average intervention effects model, the intervention effect is assumed to decrease the odds of underage drinking by 30\% on average, $\delta = \log(0.7) = -0.357$. Moreover, a nested exchangeable correlation structure is used with ICCs $(\alpha_1, \alpha_2) = (0.02, 0.01)$. From the power calculation results, power is close to 90\% given the parameters. Thus, this example further illustrates the flexibility of the SAS macro in calculating power for multi-period cluster randomized trials with different designs.
\begin{lstlisting}{sas}

%CRTFASTGEEPWR(alpha=0.05, m =%str(J(2,1,20)),  corr_type = NE,alpha1 = 0.02,
alpha2 = 0.01,intervention_effect_type=AVE, delta = -0.357,period_effect_type=CAT,
beta_period_effects =%str({0.405,-0.01,-0.01}) ,dist = binary,phi=1,
CP_size_matrix =  %str(J(2,3,30)),DesignPattern =%str({0 1 1,
                                                       0 0 0}));
\end{lstlisting}
\begin{lstlisting}{sas}
Output of EUDL design: The fast GEE power of binary outcomes with nested 
exchangeable correlation structure and (alpha1,alpha2):(0.02, 0.01) 
under average intervention effects model and delta = -0.357

T   S   clusters  df   theta   totaln    Dist   Link  stddel   zpower   tpower
3   2    40       36   0.405   3600     BINARY  LOGIT  3.2624  0.9036   0.8875        
                       -0.01
                       -0.01
                       -0.357                                                                     
\end{lstlisting}
We provide SAScodes to calculate and compare powers using GEE analysis bases on different effect sizes. In the example codes, \%CRTFASTGEEPWR is used to calculate powers under varying effect sizes, reducing the odds of underage drinking in the EDUL study by $(20\%, 25\%, 30\%, 35\%, 40\%)$ on average, leading to $\delta =log(20\%, 25\%, 30\%, 35\%, 40\%) = (-0.223,  -0.288, -0.357 , -0.431,  -0.511)$. Outputs of the SAScodes are attached below the codes.

\begin{lstlisting}{sas}
%macro  multi_effectsizes(effectsizes);
   %local index value;
   %do index = 1 %to %sysfunc(countw(&effectsizes,%str( )));
   %let value =%scan(&effectsizes,&index,%str( ));
	 %CRTFASTGEEPWR(alpha=0.05, m =%str(J(2,1,20)), corr_type = NE,alpha1 = 0.02,
	 alpha2 = 0.01 ,intervention_effect_type=AVE,  delta = &value, 
	 period_effect_type=CAT, beta_period_effects =%str({0.405,-0.01,-0.01}) ,
	 dist = binary,phi=1,CP_size_matrix =  %str(J(2,3,30)),				       		
	 DesignPattern =   %str({0 1 1,
	                         0 0 0}));
   %end; 
%mend;
%multi_effectsizes(-0.223  -0.288 -0.357  -0.431  -0.511);          
\end{lstlisting}
\newpage
\begin{lstlisting}{sas}
Outputs: The fast GEE power of binary outcomes with nested exchangeable 
correlation structure and (alpha1,alpha2):(0.02, 0.01) under average 
intervention effects model and delta = -0.223

T    S    clusters   df   theta    totaln  Dist   Link   stddel  zpower   tpower        
3    2     40        36   0.405    3600   BINARY  LOGIT  2.0482  0.5352   0.5080
                          -0.01	 	 	 	 	 	 
                          -0.01	 	 	 	 	 	 
                          -0.223	 	 	 
                        
The fast GEE power of binary outcomes with nested exchangeable 
correlation structure and (alpha1,alpha2):(0.02, 0.01) under average 
intervention effects model and delta = -0.288

T    S    clusters   df   theta    totaln  Dist   Link   stddel  zpower   tpower        
3    2     40        36   0.405    3600   BINARY  LOGIT  2.6395  0.7516   0.7276
                          -0.01	 	 	 	 	 	 
                          -0.01 	 	 	 	
                          -0.288	 
                        
The fast GEE power of binary outcomes with nested exchangeable 
correlation structure and (alpha1,alpha2):(0.02, 0.01) under average 
intervention effects model and delta = -0.357

T    S    clusters   df   theta    totaln  Dist   Link   stddel  zpower   tpower        
3    2     40        36   0.405    3600   BINARY  LOGIT  3.2624  0.9036   0.8875
                          -0.01	 	 	 	 	 	 
                          -0.01	 	 	 	 	 	 
                          -0.357	 
 	 	 	 	        
The fast GEE power of binary outcomes with nested exchangeable 
correlation structure and (alpha1,alpha2):(0.02, 0.01) under average 
intervention effects model and delta = -0.431


T    S    clusters   df   theta    totaln  Dist   Link   stddel  zpower   tpower        
3    2     40        36   0.405    3600   BINARY  LOGIT  3.9239  0.9752   0.9670
                          -0.01	 	 	 	 	 	 
                          -0.01	 	 	 	 	 	 
                          -0.431	 
                        
The fast GEE power of binary outcomes with nested exchangeable 
correlation structure and (alpha1,alpha2):(0.02, 0.01) under average 
intervention effects model and delta = -0.511

T    S    clusters   df   theta    totaln  Dist   Link   stddel  zpower   tpower        
3    2     40        36   0.405    3600   BINARY  LOGIT  4.6296  0.9962   0.9933
                          -0.01	 	 	 	 	 	 
                          -0.01	 	 	 	 	 	 
                          -0.511	 	 	 	 	 	 
                                                    
\end{lstlisting}

\end{appendix}
\newpage
%% -----------------------------------------------------------------------------

\end{document}